\documentclass[pra,twocolumn,superscriptaddress,showpacs,nofootinbibfloatfix,amsmath,amsfonts,amssymb]{revtex4-1}%
\usepackage{amsmath,amsfonts,amssymb,color}
\usepackage{amsthm}
\usepackage{leftidx}
\usepackage{graphicx}
\usepackage{xcolor}
\usepackage{dcolumn}
\usepackage{bm}
\usepackage{epstopdf}
\usepackage{epsfig}
\usepackage{environ}
\usepackage{pdfcomment}

\usepackage{float}
\usepackage[T1]{fontenc}
\usepackage[latin9]{inputenc}
\usepackage{setspace}
\usepackage{esint}

\begin{document}




\title{Dual topological characterization of non-Hermitian Floquet phases}

\author{Longwen Zhou}
\email{zhoulw13@u.nus.edu}
\affiliation{%
	Department of Physics, College of Information Science and Engineering, Ocean University of China, Qingdao, China 266100
}
\author{Yongjian Gu}
\affiliation{%
	Department of Physics, College of Information Science and Engineering, Ocean University of China, Qingdao, China 266100
}
\author{Jiangbin Gong}%
\email{phygj@nus.edu.sg}
\affiliation{%
	Department of Physics, National University of Singapore, Singapore 117543
}

\date{\today}

\begin{abstract}
Non-Hermiticity is expected to add far more physical features to the already rich Floquet topological phases of matter. Nevertheless, a systematic approach to characterize non-Hermitian Floquet topological matter is still lacking.  In this work we introduce a dual scheme to characterize the topology of non-Hermitian Floquet systems in momentum space and in real space, using a piecewise quenched nonreciprocal Su-Schrieffer-Heeger model for our case studies.  Under the periodic boundary condition, topological phases are characterized by a pair of experimentally accessible winding numbers that make jumps between integers and half-integers.  Under the open boundary condition, a Floquet version of the so-called open boundary winding number is found to be integers and can predict the number of pairs of zero and $\pi$ Floquet edge modes coexisting with the non-Hermitian skin effect. Our results indicate that a dual characterization of non-Hermitian Floquet topological matter is necessary and also feasible because the formidable task of constructing the celebrated generalized Brillouin zone for non-Hermitian Floquet systems with multiple hopping length scales can be avoided.  This work hence paves a way for further studies of non-Hermitian physics in non-equilibrium systems.   
\end{abstract}

\pacs{}
\keywords{}
\maketitle

\textit{Introduction.--}Floquet topological phases, as created by time-periodic modulations, have been an experimental reality in both synthetic metamaterials \cite{Rechtsman2013,Szameit2017,Cheng2019,Aidelsburger2020} and actual condensed-matter systems \cite{Wang2013Exp,Mciver2020Exp}.  One genuinely promising feature of such nonequilibrium topological phases is that they may accommodate an arbitrary number of topological edge modes \cite{HFTP8}, e.g., the coexistence of many chiral edge modes to enhance robust transport~\cite{HFTP6,HFTP7}, and on-demand generation of multiple dispersionless edge modes \cite{HFTP1,HFTP3,LWZFTI1} for encoding and processing quantum information \cite{RadityaPRL08,RadityaPRB08}.  To further explore far-reaching possibilities offered by Floquet topological matter, it is timely and potentially fruitful to introduce non-Hermiticity to periodically driven lattice systems. 
 
The interplay between periodic driving and non-Hermiticity is expected to be rich \cite{WangPRA2018,WangPRA2019,LWZNH1} and has already led to some encouraging findings~\cite{LWZNH2,LWZNH3,LWZNH4,NHFTI1,NHFTI2,NHFTI3,NHFTI4,LWZNH5,LWZNH6,NHFTSC1,NHFTSC2,NHFTSM1,NHQW1,NHQW2,NHQW3,NHQW4,CHCP2020}. 
Two aspects of non-Hermitian Floquet matter are worthy of special attention.   First, the exceptional topology~\cite{rev1,rev2,rev3} in the complex Floquet band structure can potentially create even richer topological phases absent in Hermitian cases.  Independent of other topological aspects of Floquet bands, characterizing and experimentally detecting Floquet exceptional topology are of general interest. Second, the so-called non-Hermitian skin effect (NHSE)~\cite{Wang1,Wang2,NHSSH0,NHSE3,NHSE4,NHSE26,skinnew1,skinnew2}, which corresponds to the pile up of bulk states at the edges of a non-Hermitian lattice, must also be well addressed for a topological characterization aiming at predicting the emergence of many topological edge modes, localized not because of NHSE, but topological localization.   Remarkably, the very main reason of why Floquet topological phases can be so rich, namely, the emergent/effective hoppings across different and extended hopping ranges \cite{HFTP1} in the same system, presents a severe challenge  in analyzing the NHSE with the celebrated generalized Brillouin zone (GBZ) treatment~\cite{Wang2,NHSSH0,NHSE3,NHSE4,NHSE26,skinnew1,skinnew2,skinnew3}. That is, the GBZ would be too hard to be computationally constructed in Floquet systems with the coexistence of many different length scales~\cite{FNHSE1}.

Here we propose a dual topological characterization scheme to investigate a representative and simple class of non-Hermitian Floquet matter protected by chiral symmetry, in both momentum space and real space. Under the periodic boundary condition (PBC),  the exceptional topology in the Floquet bands yields a phase diagram characterized by two species of winding numbers depicting Floquet effective Hamiltonians in two time frames, with the winding numbers being tunable without  a bound and alternating between integers and half-integers. These intriguing topological phases correspond to gap closing and reopening at eigenphase zero or $\pi$ and can be directly probed in experiments. Under the open boundary condition (OBC),  more complications of topological characterization arise because an unlimited number of topological edge modes at eigenphases zero and $\pi$ can coexsit with NHSEs. We propose a Floquet version of the so-called open-boundary winding numbers~(OBWNs)~\cite{NHSSH0} in real space.  Two types of OBWNs are advocated, being always integers,  and they precisely match the number of pairs of the two different types of Floquet edge modes. The OBWNs jump only when the spectral gap under OBC closes and reopens. As further elaborated below, the topological characterization in momentum- and that in real-space are different but also related, thus indicating the necessity of a dual approach for a complete picture of non-Hermitian Floquet topological phases.  It is also tempting to view the seen differences as evidence of a must breakdown of the old concept of bulk-edge correspondence.

\begin{figure}
	\begin{centering}
		\includegraphics[scale=0.42]{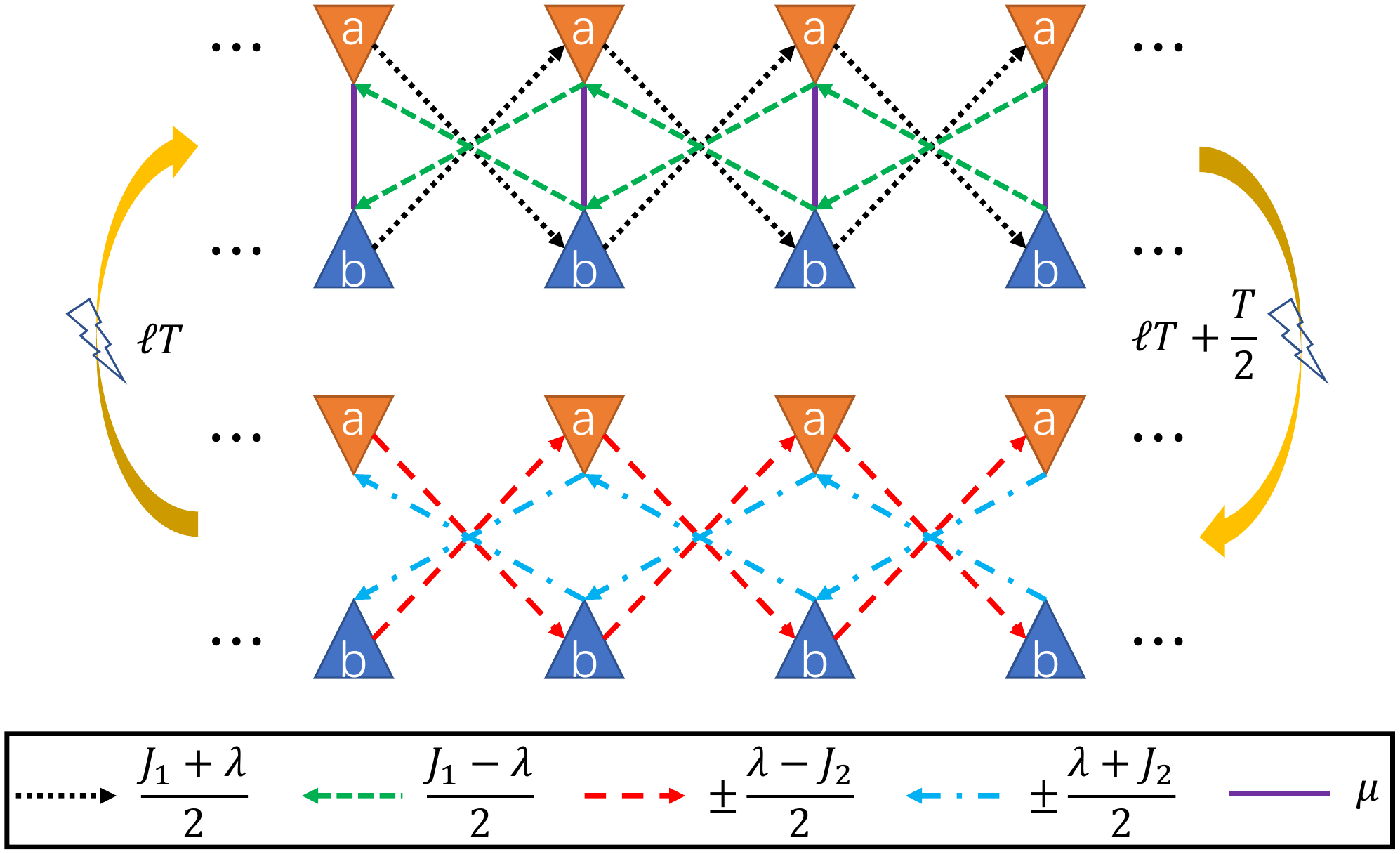}
		\par\end{centering}
	\caption{Schematic illustration of the periodically quenched NHSSH model.
		Each unit cell contains two sublattices, which are coupled
		by the intracell hopping amplitude $\mu$. The intercell
		hopping amplitudes in the first and second halves of the driving period
		are $\frac{J_{1}\pm\lambda}{2}$ and $\pm\frac{\lambda\pm J_{2}}{2}$.
		The lightning symbols denote quenches applied in the middle/end
		of each driving period, after which the lattice is switched from the
		configuration of the upper/lower to the lower/upper array of the
		figure.\label{fig:Sketch}}
\end{figure}

\textit{Model.--}To make our theoretical considerations more explicit, we start with a non-Hermitian Su-Schrieffer-Heeger (NHSSH) model~\cite{NHSSH0} under periodic quenches. In momentum space, the Hamiltonian of the model takes the form
\begin{equation}
H(k,t)=\begin{cases}
H_x(k)=h_{x}(k)\sigma_{x} & t\in\left[\ell T,\ell T+\frac{T}{2}\right)\\
H_y(k)=h_{y}(k)\sigma_{y} & t\in\left[\ell T+\frac{T}{2},\ell T+T\right)
\end{cases}.\label{eq:Hkt}
\end{equation}
Here the quasimomentum $k\in[-\pi,\pi)$, $\ell\in\mathbb{Z}$, $T$ is the driving period, and $\sigma_{x,y,z}$
are Pauli matrices acting on the sublattice degrees of freedom. The components of the Hamiltonian are given by $h_{x}(k)=\mu+J_{1}\cos k+i\lambda\sin k$ and $h_{y}(k)=J_{2}\sin k+i\lambda\cos k$. $\mu$ and $\frac{J_{1}+J_{2}}{2}$ are the intracell and intercell
hopping amplitudes of the SSH model. The non-Hermiticity is introduced by asymmetric hoppings $\frac{J_{1}-J_{2}}{2}\pm i\frac{\lambda}{2}$
between the two sublattices. An illustration of the model is given in Fig.~\ref{fig:Sketch}. 

We set $\hbar=1$ throughout and the driving period $T=2$. Following Eq.~(\ref{eq:Hkt}), the Floquet operator depicting the time evolution is $U(k)=e^{-ih_{y}(k)\sigma_{y}}e^{-ih_{x}(k)\sigma_{x}}$. Referring to the established topological characterization of 1D Floquet systems~\cite{AsbothSTF1,AsbothSTF2,LWZFTI1}, we introduce a pair of symmetric time frames, in which $U(k)$ take the forms
$U_{1}(k)=e^{-i\frac{h_{x}(k)}{2}\sigma_{x}}e^{-ih_{y}(k)\sigma_{y}}e^{-i\frac{h_{x}(k)}{2}\sigma_{x}}=e^{-iH_{1}(k)}$ and $U_{2}(k)=e^{-i\frac{h_{y}(k)}{2}\sigma_{y}}e^{-ih_{x}(k)\sigma_{x}}e^{-i\frac{h_{y}(k)}{2}\sigma_{y}}=e^{-iH_{2}(k)}$.
Since $U(k)$ and $U_{1,2}(k)$ are related by similarity transformations, they share the same Floquet eigenphase spectrum $E(k)$, which can be obtained by
solving $H_{\alpha}(k)|\psi_{\alpha}^{\pm}(k)\rangle=\pm E(k)|\psi_{\alpha}^{\pm}(k)\rangle$
for $\alpha=1,2$~\cite{Note1}. 
With Taylor expansions of $e^{-i\frac{h_{x,y}(k)}{2}\sigma_{x,y}}$, $e^{-ih_{x,y}(k)\sigma_{x,y}}$, and combining the resulting terms, we find the effective Hamiltonians
\begin{equation}
H_{\alpha}(k)=h_{\alpha x}(k)\sigma_{x}+h_{\alpha y}(k)\sigma_{y},\quad\alpha=1,2.\label{eq:Hak}
\end{equation}
$H_{\alpha}(k)$ possesses the chiral (sublattice) symmetry ${\cal S}=\sigma_{z}$,
time-reversal symmetry ${\cal T}=\sigma_{0}$ and particle-hole symmetry
${\cal C}=\sigma_{z}$, i.e., ${\cal S}H_{\alpha}(k){\cal S}=-H_{\alpha}(k)$,
${\cal T}H_{\alpha}^{*}(k){\cal T}^{-1}=H_{\alpha}(-k)$, and ${\cal C}H_{\alpha}^{*}(k){\cal C}^{-1}=-H_{\alpha}(-k)$, where $\sigma_{0}$ denotes the $2\times2$ identity
matrix. 
The system under study hence belongs to the symmetry class BDI~\cite{NHClass1,NHClass2,NHClass3}.
The symmetry ${\cal S}$ ensures that the eigenvalues of $H_\alpha(k)$ appear
in positive-negative pairs on the complex plane, yielding the topological protection of Floquet edge modes at $E=0,\pi$. In addition, $H_{\alpha}(k)$ also lacks the inversion symmetry, indicating the existence of NHSEs~\cite{NHClass2}.


\textit{Momentum-space characterization.--}Since $H_{\alpha}(k)$ possesses the chiral symmetry, we proceed to use the following winding number $w_{\alpha}$~\cite{LWZNH1}
\begin{equation}
w_{\alpha}=\int_{-\pi}^{\pi}\frac{dk}{2\pi}\partial_{k}\phi_{\alpha}(k), \quad\alpha=1,2, \label{eq:Wa}
\end{equation}
where the winding angle $\phi_{\alpha}(k)\equiv\arctan[h_{\alpha y}(k)/h_{\alpha x}(k)]$. $w_{\alpha}$ counts the number of times the angle $\phi_{\alpha}(k)$ changes over $2\pi$ as the quasimomentum $k$ sweeps across
the first BZ. Thus, this topological invariant is based entirely from the momentum-space effective Hamiltonian in two different time frames.
Notably, $w_{\alpha}$ is highly nontrivial because $h_{\alpha x}(k)$ and $h_{\alpha y}(k)$ are complex functions. Furthermore, the imaginary
part of $\phi_{\alpha}(k)$ has no contribution to the integral
over $k$, and $w_{\alpha}$ is hence real~\cite{LWZNH3,DWN1}. More importantly, except for some special initial states,  $w_{\alpha}$ thus defined can be measured dynamically
by averaging some spin textures over a sufficiently long time~\cite{LWZNH1,LWZNH3}.

Back to the original time frame, we can now introduce two species of invariants to characterize the bulk topological properties. With the winding
numbers $(w_{1},w_{2})$, we define topological invariants 
\begin{equation}
w_{0}=\frac{w_{1}+w_{2}}{2},\qquad w_{\pi}=\frac{w_{1}-w_{2}}{2}.\label{eq:W0P}
\end{equation}
As confirmed below, $(w_{0},w_{\pi})$ are respectively protected by Floquet band gaps around eigenphases zero and $\pi$.
Interestingly, the system parameters chosen in previous studies~\cite{LWZNH1,LWZNH2,LWZNH3,LWZNH4} happened to guarantee that always integer values of $(w_{0},w_{\pi})$  can be obtained. However, we discover that in more general situations,  $w_{0}$ or $w_{\pi}$ can take half-integer values, small or large.  In particular, half-integer $w_\pi$ indicates the emergence of exceptional topology due to the gap closing at $E=\pi$, which is absent in static systems. Such half-integer windings should not be connected with the number of possible Floquet edge modes (because there cannot be a half pair of topological edge modes for the symmetry class under consideration). Nevertheless, since $(w_{0},w_{\pi})$ can be measured from dynamical spin textures \cite{LWZNH1,LWZNH3} and are robust to perturbations that preserve the chiral symmetry, they do present together a momentum-space topological characterization.

\begin{figure}
	\begin{centering}
		\includegraphics[scale=0.49]{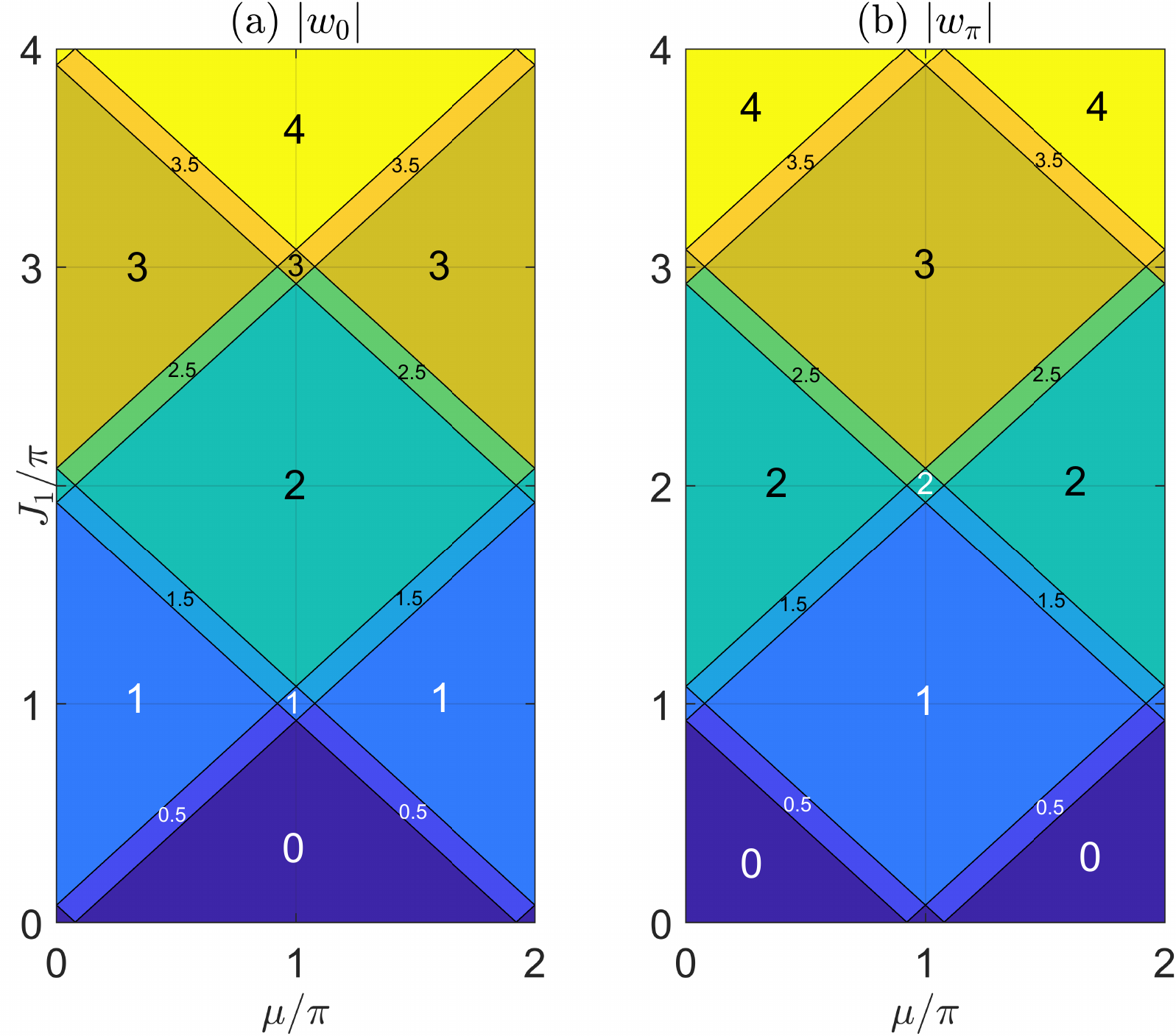}
		\par\end{centering}
	\caption{Bulk topological phase diagram of the periodically quenched NHSSH
		model versus hopping parameters $\mu$ and $J_{1}$. Other system
		parameters are $(J_{2},\lambda)=(0.5\pi,0.25)$. Each region
		with a uniform color denotes a topological phase characterized
		by the winding numbers $(w_{0},w_{\pi})$. The solid lines between
		different regions are boundaries between distinct non-Hermitian
		Floquet topological phases. The values of $w_{0}$ and $w_{\pi}$ for each phase are denoted explicitly in panels (a) and (b).\label{fig:PhsDiag1}}
\end{figure}
As a typical case, we present the topological phase diagram of the periodically quenched NHSSH model versus the hopping parameters in Fig.~\ref{fig:PhsDiag1}. The values of $w_{0}$ and $w_{\pi}$
are obtained from Eqs.~(\ref{eq:Wa})-(\ref{eq:W0P}) and marked explicitly in each region of the left and right panels. Different non-Hermitian Floquet topological phases are distinguished
by their colors.
It is seen that with the change of $J_{1}$ and $\mu$, the system undergoes a series of topological phase transitions, which are accompanied
by quantized or half-quantized jumps of $w_{0}$ or $w_{\pi}$. Furthermore,
with the increase of $J_{1}$, a monotonous increase of
the values of $w_{0}$ or $w_{\pi}$ across each transition is observed, yielding
non-Hermitian Floquet states characterized by large integers
or large half-integers for both species of winding numbers.  These intriguing phases  are unique to non-Hermitian Floquet systems~\cite{Note1}.  

\begin{figure}
	\begin{centering}
		\includegraphics[scale=0.49]{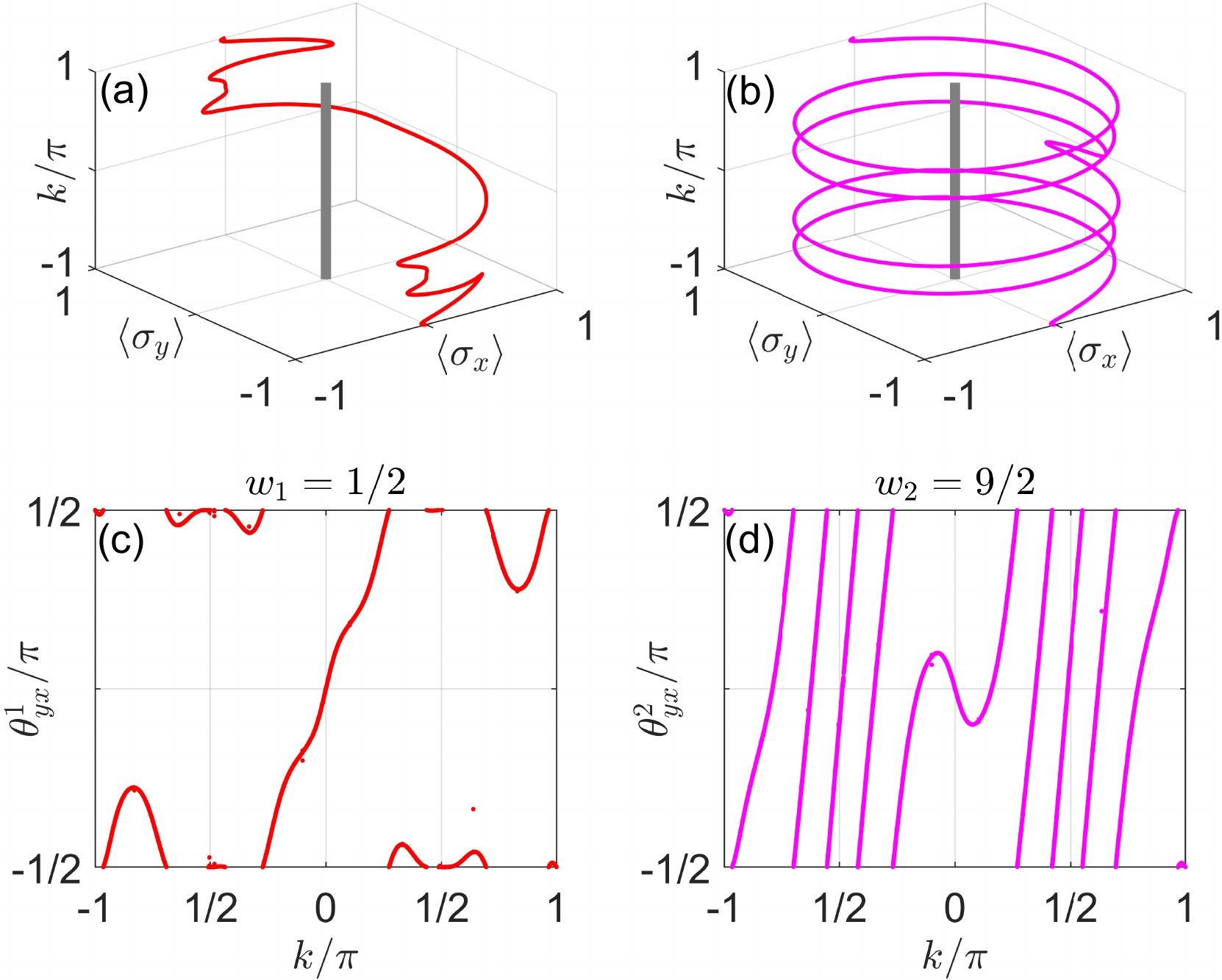}
		\par\end{centering}
	\caption{Spin textures and dynamic winding angles of the periodically quenched NHSSH model in time frames $\alpha=1$ [panels (a,c)] and $\alpha=2$ [panels (b,d)]. System parameters are $(J_1,J_2,\mu,\lambda)=(2.4\pi,0.5\pi,0.4\pi,0.25)$, and evolutions are averaged over $500$ driving peirods to generate the winding angles $\theta^{1,2}_{yx}$ in panels (c,d). In the panel (a) [(b)], the red (magenta) points denote $(\langle\sigma_x\rangle,\langle\sigma_y\rangle)$ in the first (second) time frame. The grey solid line denotes the origin of $\langle\sigma_x\rangle$-$\langle\sigma_y\rangle$ plane. In panel (c) [(d)], the red (magenta) points correspond to the dynamic winding angles $\theta^{1}_{yx}$~($\theta^{2}_{yx}$) in time frame $1$~($2$)~\cite{LWZNH3}. The values of $(w_1,w_2)$ are also shown in panels (c,d).\label{fig:Spintext2}}
\end{figure}

To digest the physical meanings of the half-integer winding numbers, we present the long-time averaged spin textures and dynamic winding numbers~\cite{LWZNH3} for a typical situations in Fig.~\ref{fig:Spintext2}, where the panels (a) and (b) show the trajectors of spin vector $(\langle\sigma_x\rangle,\langle\sigma_y\rangle)$ versus the quasimomentum $k$ in the time frame $\alpha=1$ and $2$. The average $\langle\cdots\rangle$ is taken with respect to the right eigenvector $|\psi_{\alpha}^{+}(k)\rangle$ of $H_\alpha(k)$. The grey thick lines highlight the origin of the $\langle\sigma_x\rangle$-$\langle\sigma_y\rangle$ plane, which satisfy the equation $(\langle\sigma_x\rangle,\langle\sigma_y\rangle)=0$ at all $k\in[-\pi,\pi)$. In Figs.~\ref{fig:Spintext2}(a,b), we see that the projection of $(\langle\sigma_x\rangle,\langle\sigma_y\rangle)$ on $\langle\sigma_x\rangle$-$\langle\sigma_y\rangle$ plane contains an integer plus a half circle, which indicates the presence of half-integer winding numbers. For example, starting at $(\langle\sigma_x\rangle,\langle\sigma_y\rangle)=(0,-1)$, the vector $(\langle\sigma_x\rangle,\langle\sigma_y\rangle)$  rotates counterclockwise around the origin over $4.5$ cycles, ending at $(\langle\sigma_x\rangle,\langle\sigma_y\rangle)=(0,1)$ when $k$ sweeps from $-\pi$ to $\pi$, as shown in Fig.~\ref{fig:Spintext2}(b). These half-integer windings are caught by the winding angles of dynamic spin textures $\theta^{1,2}_{yx}$, as shown in panels (c,d) of Fig.~\ref{fig:Spintext2}~(see Ref.~\cite{LWZNH3} for the definition and calculation of the dynamic winding angles), where the net increments of $\theta^{1,2}_{yx}$ across the first BZ are odd-integer multiples of $\pi$, yielding half-quantized integers after being divided by $2\pi$. In Ref.~\cite{LWZNH3}, it was proven that $(\theta^{1}_{yx}/2\pi,\theta^{2}_{yx}/2\pi)$ are equal to $(w_1,w_2)$ defined in Eq.~(\ref{eq:Wa}). Therefore, if $(\theta^{1}_{yx}+\theta^{2}_{yx})/(2\pi)$ or $(\theta^{1}_{yx}-\theta^{2}_{yx})/(2\pi)$ happens to be an odd integer, we obtain a half-quantized invariant $w_0$ or $w_\pi$ according to Eq.~(\ref{eq:W0P}). Thus, the half-integer quantization of $(w_0,w_\pi)$ can also be dynamically extracted from time-averaged spin textures~\cite{Note1}.

Qualitatively, the half windings of $w_1$ and $w_2$ may be
 be traced back to the branch switch of the two Floquet bands when $k$ varies from $-\pi$ to $\pi$~\cite{NHSE26}. Together with the above-mentioned symmetry between the upper and lower complex plane of the Floquet spectrum, the band switch (braiding) indicates that there is necessarily windings of the Floquet spectral flow on the complex plane, thus signaling the existence of NHSE \cite{NHSE3,NHSE4,NHSE26,skinnew1, skinnew1,skinnew2}. To treat the possible coexistence of many Floquet edge modes with NHSE, we next move on to real-space characterization. 
 


\textit{Real-space topological characterization.--}The above-obtained large winding numbers in momentum space  already indicate the existence of multiple scales of hopping in the Floquet effective Hamiltonians. In such situations, construction of a GBZ to treat NHSE is not practical. This motivates us to extend the OBWN previously for static non-Hermitian systems~\cite{NHSSH0} to non-Hermitian Floquet lattices.

We first define the ${\cal Q}$-matrix~\cite{Note1} in a time frame $\alpha$ as ${\cal Q}_{\alpha}=\sum_{n}(|\psi_{\alpha n}^{+}\rangle\langle\tilde{\psi}_{\alpha n}^{+}|-|\psi_{\alpha n}^{-}\rangle\langle\tilde{\psi}_{\alpha n}^{-}|)$. The right (left) Floquet
eigenvectors $|\psi_{\alpha n}^{\pm}\rangle$ ($\langle\tilde{\psi}_{\alpha n}^{\pm}|$)
satisfy the eigenvalue equations $U_{\alpha}|\psi_{\alpha n}^{\pm}\rangle=e^{-i(\pm E_{n})}|\psi_{\alpha n}^{\pm}\rangle$
[$\langle\tilde{\psi}_{\alpha n}^{\pm}|U_{\alpha}=\langle\tilde{\psi}_{\alpha n}^{\pm}|e^{-i(\pm E_{n})}$],
with $\pm E_{n}$ being the eigenphases. $U_{\alpha}$ is given by the real-space representation
of $U_{\alpha}(k)$. With ${\cal Q}_{\alpha}$, one can construct the Floquet OBWN as
\begin{equation}
\nu_{\alpha}=\frac{1}{L_{\rm B}}{\rm Tr}_{\rm B}({\cal S}{\cal Q}_{\alpha}[{\cal Q}_{\alpha},{\cal N}]).\label{eq:NUa}
\end{equation}
${\cal S}=\mathbb{I}_{N\times N}\otimes\sigma_{z}$ is the chiral (sublattice) symmetry operator. $\mathbb{I}_{N\times N}$ is an $N\times N$ identity matrix and $N$
is the total number of unit cells. $L_{\rm B}$ and ${\rm Tr}_{\rm B}$ share
the same physical meanings as in the static version of OBWN~\cite{NHSSH0}. That is,
with the system decomposed into a bulk region and two edge regions around the left
and right boundaries, the trace ${\rm Tr}_{\rm B}$ is taken over the bulk region,
which contains $L_{\rm B}$ lattice sites. Further, for a lattice of $L$ sites, the length of each
edge region is $L_{\rm E}=(L-L_{\rm B})/2$. Though we make no attempt to construct a GBZ (which cannot be done),
$\nu_\alpha$ thus defined should, just as expected from the static case~\cite{NHSSH0}, essentially yields a winding number of the effective Hamiltonian $H_\alpha(k)$ in the $\alpha$th time frame along the underlying GBZ.
Finally, as one essential step in our treatment and in analogy to the definition of $(w_{0},w_{\pi})$ in Eq.~(\ref{eq:W0P}), we define two types of OBWNs
of a 1D non-Hermitian Floquet system as
\begin{equation}
\nu_{0}=\frac{\nu_{1}+\nu_{2}}{2},\qquad\nu_{\pi}=\frac{\nu_{1}-\nu_{2}}{2}.\label{eq:NU0P}
\end{equation}
$(\nu_{0},\nu_{\pi})\in{\mathbb Z}\times{\mathbb Z}$ defined above serve as two new topological invariants arising from our real-space characterization for Floquet systems.  We present below compelling evidence that this OBC characterization works properly  because they can predict the numbers of topologically protected modes at eigenphases zero and $\pi$, denoted as $(n_{0},n_{\pi})$, through the bulk-edge correspondence relations $(n_{0},n_{\pi})=2(|\nu_{0}|,|\nu_{\pi}|)$.

\begin{figure}
	\begin{centering}
		\includegraphics[scale=0.48]{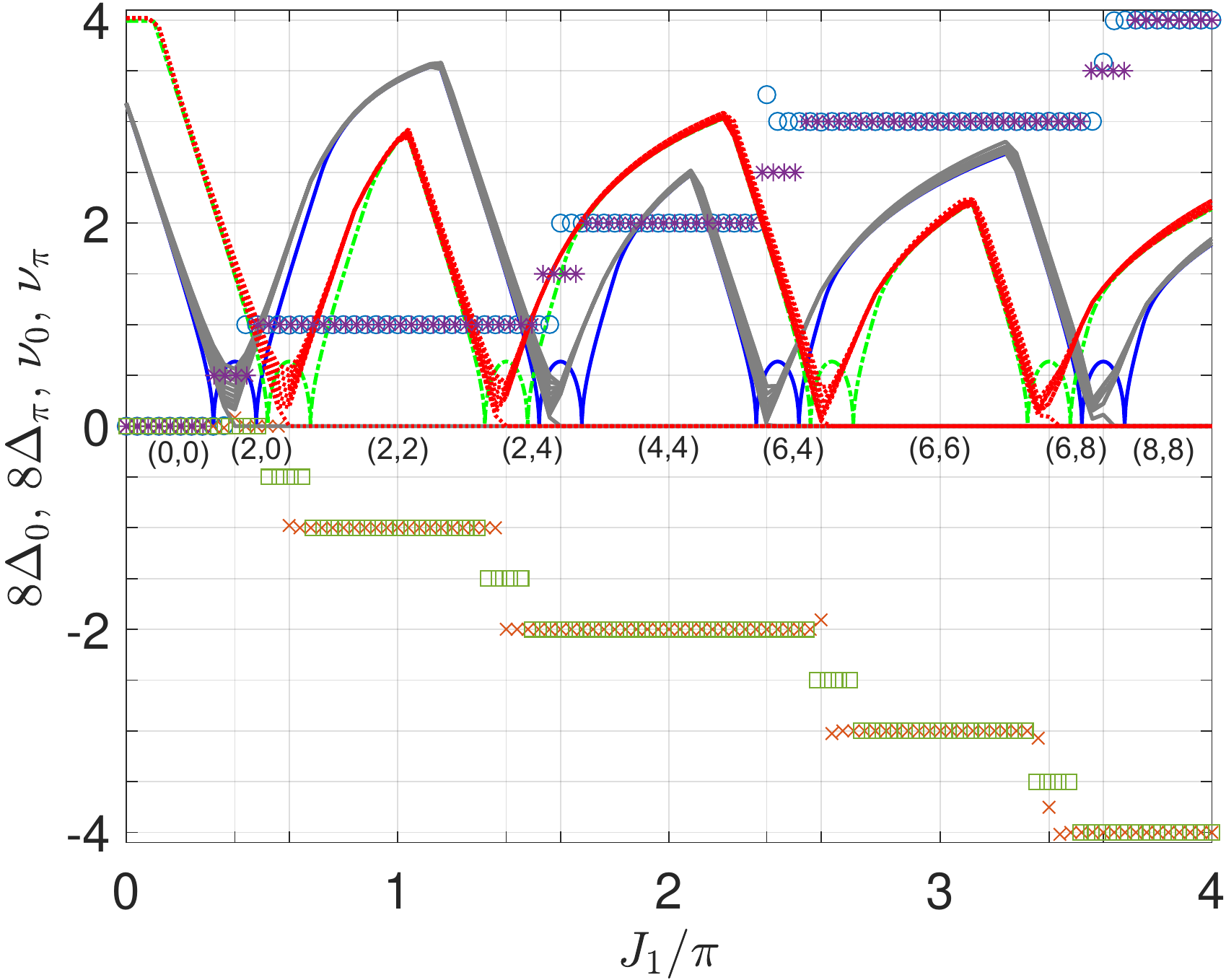}
		\par\end{centering}
	\caption{Gap functions $\Delta_{0}$ (blue and gray solid lines), $\Delta_{\pi}$ (red and green dotted lines), OBWNs $\nu_{0}$ (circles), $\nu_{\pi}$ (crosses), and PBC winding numbers $w_0$ (stars), $w_\pi$ (squares) of the model. System parameters are $(\mu,J_2,\lambda)=(0.4\pi,0.5\pi,0.25)$. Phase transitions under OBC happen at $J_{1}=(0.4\pi,0.6\pi,1.4\pi,1.6\pi,2.4\pi,2.6\pi,3.4\pi,3.6\pi)$, denoted by the ticks along the horizontal axis. Only the first twenty smallest values of $(\Delta_{0},\Delta_{\pi})$	under OBC are shown for clear illustrations. The numbers of zero and $\pi$ Floquet edge modes are denoted below the horizontal axis.\label{fig:BBC1}}
\end{figure}

Let us now compare the Floquet eigenphase spectrum of the periodically quenched NHSSH model under
the PBC and OBC. To reveal the gap closing-reopening points clearly, we introduce gap functions $\Delta_{0}=|E|/\pi$ and $\Delta_{\pi}=\sqrt{(|{\rm Re}E|-\pi)^{2}+({\rm Im}E)^{2}}/\pi$. It is clear that the Floquet spectrum become gapless at $E=0$
($E=\pi$) if $\Delta_{0}=0$ ($\Delta_{\pi}=0$), where a phase transition occurs. In Fig.~\ref{fig:BBC1}, we show $(\Delta_{0},\Delta_{\pi})$ of our model versus the hopping amplitude $J_{1}$ under both
the PBC and OBC in a lattice of $L=400$ sites. The
spectrum under PBC~(in blue solid and green dotted lines) and OBC (in gray solid and red dotted lines) are expectedly similar in regimes far from gap closing points but clearly different near the gapless points. For example,
at $J_{1}=0.4\pi$, one sees a phase transition with
$\Delta_{0}=0$ under the OBC, after which a pair of edge modes with $E=0$
emerges. However, the spectrum under PBC suggests
two consecutive transitions at $J_{1}<0.4\pi$ and $J_{1}>0.4\pi$.
In between, there is a bulk topological phase with winding number
$w_{0}=1/2$ according to Eqs.~(\ref{eq:Wa})-(\ref{eq:W0P}). Fig.~\ref{fig:BBC1} presents many other similar regimes where the momentum-space topological invariants differ from the OBC winding numbers by $1/2$.
Such a clear distinction between the Floquet spectrum  under
PBC and OBC indicates the presence of NHSEs and breakdown of
the bulk-edge correspondence~\cite{Note1}. 

We next compute $(\nu_{0},\nu_{\pi})$
following Eqs.~(\ref{eq:NUa})-(\ref{eq:NU0P}). The results are also presented in Fig.~\ref{fig:BBC1}, where the numbers of zero and $\pi$ Floquet edge modes are denoted. We see that the $(\nu_{0},\nu_{\pi})$ take integer
values within each non-Hermitian Floquet topological phase, and undergo
quantized jumps when $J_{1}$ is tuned
through a topological phase transition point, where we have  $\Delta_{0}=0$
or $\Delta_{\pi}=0$ under OBC.  Within each topological phase, $(\nu_{0},\nu_{\pi})$ correctly count the numbers of zero and $\pi$ edge modes, thus verifying the bulk-edge correspondence of our system albeit the
existence of NHSEs. Besides, with the increase of $J_{1}$, almost monotonic increases in $(\nu_{0},\nu_{\pi})$ and in the numbers of edge modes $(n_{0},n_{\pi})$ are observed. This verifies again the enormous potential of Floquet engineering in realizing non-Hermitian Floquet states of matter with in principle unbounded topological invariants available and hence as many topological edge modes as we wish.

\textit{Summary.--}We have introduced a powerful dual scheme to characterize non-Hermitian Floquet topological matter, as illustrated by a simple periodically quenched NHSSH model.   Rich non-Hermitian Floquet phases under PBC are characterized by two species of topological invariants that can be experimentally measured. The half-integer topological invariants associated with both zero and $\pi$ gaps are identified as a general feature of exceptional topology in Floquet systems. Under the OBC, topological edge modes pinned at eigenphases zero and $\pi$ can be generated in large numbers, together with NHSEs. We have found two OBWNs that can be used  to characterize the Floquet topological phases in real space, thus avoiding the formidable task of constructing a GBZ.   Interestingly, we now have two nonequivalent topological descriptions of the same Floquet system, with each of them necessary on its own right. Their differences also constitute a fascinating example of the breakdown of conventional bulk-edge correspondence (but well restored by OBWNs we proposed here). This work have thus laid a necessary and timely stage for further understanding and use of non-Hermitian Floquet phases for topology-based applications.
In future work, it would be interesting to extend our framework to higher-dimensional, disordered and many-body non-Hermitian Floquet systems, in which the concept of GBZ does not apply in general.

\textit{Acknowledgements.--}The authors acknowledge Lee Ching Hua and Li Linhu for helpful comments. L.Z. is supported by the National Natural Science Foundation of China (Grant No. 11905211), the China Postdoctoral Science Foundation (Grant No. 2019M662444), the Fundamental Research Funds for the Central Universities (Grant No. 841912009), the Young Talents Project at Ocean University of China (Grant No. 861801013196), and the Applied Research Project of Postdoctoral Fellows in Qingdao (Grant No. 861905040009). J.G. acknowledges support from Singapore National Research Foundation Grant No. NRF- NRFI2017-04 (WBS No. R-144-000-378-281). Y.G. acknowledges support from National Natural Science Foundation of China (Grant No. 61575180).


\clearpage
\newpage{}

\section*{Supplementary Material\label{sec:Sup}}

In this Supplementary Material, we provide more details about the model and methods developed in the main text. In Sec.~\ref{subsec:A}, we give the explicit expressions of the eigenphase dispersion $E(k)$ and effective Hamiltonian components $[h_{\alpha x}(k),h_{\alpha y}(k)]$ of the periodically quenched NHSSH model studied in the main text. In Sec.~\ref{subsec:B}, we provide more examples of the bulk topological phase diagrams of our model. In Sec.~\ref{subsec:C}, we present another example of the static spin textures and dynamic winding numbers of the system studied in the main text. In Sec.~\ref{subsec:D}, we give the explicit expressions of the system's Hamiltonian under the OBC. In Sec.~\ref{subsec:E}, we present the definition and calculation details of the open boundary ${\cal Q}$-matrix in static non-Hermitian systems. In Sec.~\ref{subsec:F}, we give more examples of the Floquet spectrum and OBWNs of the periodically quenched NHSSH model. In Sec.~\ref{subsec:G}, we present the profiles of bulk Floquet skin modes and edge states of our model.

\subsection{Expressions for the spectrum $E(k)$ and the components of $H_{\alpha}(k)$\label{subsec:A}}

In the main text, we obtained the effective Hamiltonian $H_{\alpha}(k)=h_{\alpha x}(k)\sigma_{x}+h_{\alpha y}(k)\sigma_{y}$ in the two time frames $\alpha=1,2$ by taking the Taylor expansion and applying the Euler formula to the Floquet operator $U_{\alpha}(k)$. The components $h_{\alpha x}(k)$ and $h_{\alpha y}(k)$ are explicitly given by:
\begin{alignat}{1}
h_{1x}(k) & =\frac{E(k)\sin[h_{x}(k)]\cos[h_{y}(k)]}{\sin[E(k)]},\label{eq:h1x}\\
h_{1y}(k) & =\frac{E(k)\sin[h_{y}(k)]}{\sin[E(k)]},\label{eq:h1y}\\
h_{2x}(k) & =\frac{E(k)\sin[h_{x}(k)]}{\sin[E(k)]},\label{eq:h2x}\\
h_{2y}(k) & =\frac{E(k)\sin[h_{y}(k)]\cos[h_{x}(k)]}{\sin[E(k)]}.\label{eq:h2y}
\end{alignat}
Here $E(k)=\arccos\{\cos[h_{x}(k)]\cos[h_{y}(k)]\}$ is the eigenphase dispersion relation. Due to the chiral symmetry, the Floquet spectrum gaps could vanish at $E(k)=0$ and $E(k)=\pi$, yielding possible topological phase transitions under the PBC as discussed in the main text. 

\subsection{More examples of the bulk topological phase diagrams\label{subsec:B}}

In the main text, we showed a typical example of the topological phase diagram of our periodically quenched NHSSH model, which is characterized by the introduced bulk winding numbers $(w_0,w_\pi)$. In this section, we support our conclusion with two more examples of the topological phase diagrams.

\begin{figure}
	\begin{centering}
		\includegraphics[scale=0.48]{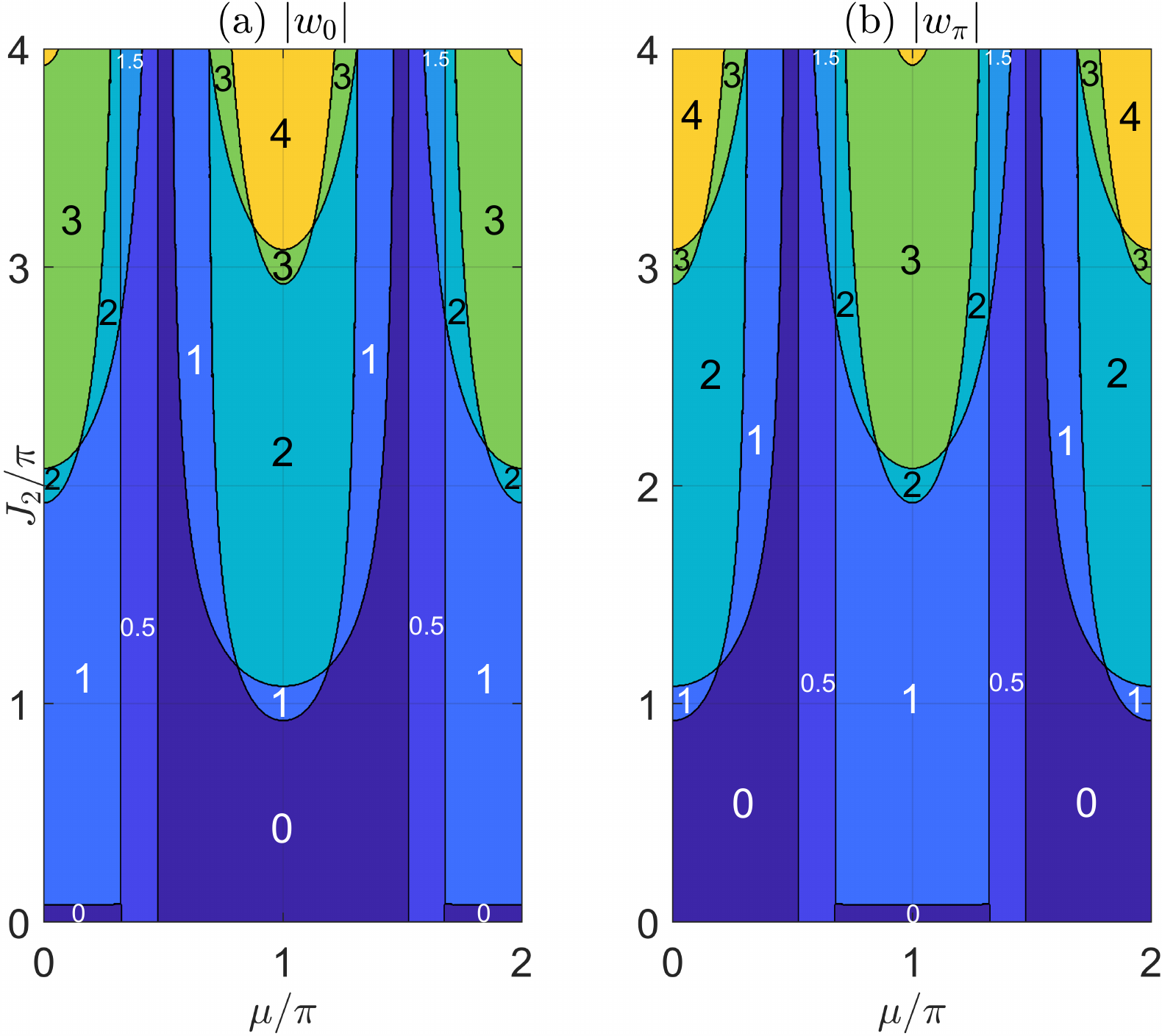}
		\par\end{centering}
	\caption{Bulk topological phase diagram of the periodically quenched NHSSH
		model with respect to the hopping amplitudes $\mu$ and $J_{2}$.
		The other system parameters are chosen to be $J_{1}=0.4\pi$ and $\lambda=0.25$.
		In panels (a) and (b), each patch with a uniform color refers to a
		non-Hermitian Floquet phase of the model, whose winding numbers $w_{0}$
		and $w_{\pi}$ are denoted explicitly therein. The solid lines separating
		different regions represent the boundaries of different topological
		phases, where the spectrum gap closes at the quasienergy zero {[}in
		panel (a){]} and $\pi$ {[}in panel (b){]}.\label{fig:PhsDiag2}}
\end{figure}

\begin{figure}[h]
	\centering
	\includegraphics[scale=0.40, trim=0cm 1.5cm 0cm 1.5cm, clip=true]{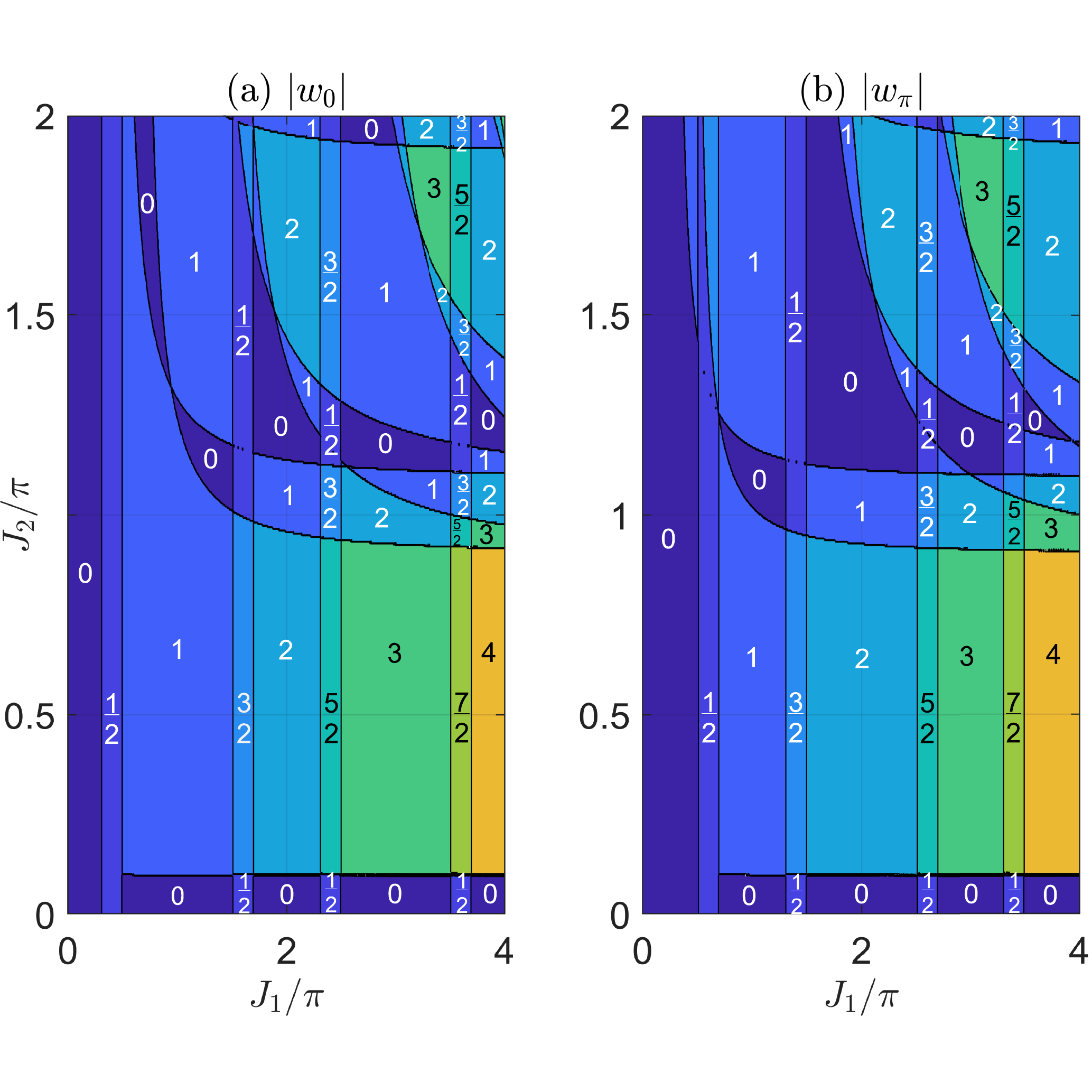}
	\caption{Bulk topological phase diagram of the periodically quenched NHSSH
		model versus the hopping amplitudes $J_{1}$ and $J_{2}$. The other
		system parameters are elected as $\mu=0.4\pi$ and $\lambda=0.3$.
		In both panels, each regime with a even color corresponds to a non-Hermitian
		Floquet topological phase of the system. The values of winding numbers $w_{0}$
		and $w_{\pi}$ for each Floquet topological phase are
		presented explicitly in panels (a) and (b), respectively. The solid
		lines separating different regions are the boundaries between different
		topological phases, across which the winding numbers $w_{0}$ {[}in
		panel (a){]} and $w_{\pi}$ {[}in panel (b){]} take quantized or half-quantized
		jumps.\label{fig:PhsDiag3}}
\end{figure}

In the first case, we present the phase diagram with respect to hopping
parameters $\mu$ and $J_{2}$ in Fig.~\ref{fig:PhsDiag2}. The other
system parameters are chosen to be $J_{1}=0.4\pi$ and $\lambda=0.25$.
Each region with a uniform color in Fig.~\ref{fig:PhsDiag2} corresponds
to a Floquet topological phase of the periodically quenched NHSSH
model, whose winding numbers $w_{0}$ and $w_{\pi}$ are denoted explicitly
in the left and right panels. We observe rich non-Hermitian
Floquet topological phases and phase transitions induced by the change
of system parameters. Moreover, the winding numbers $(w_{0},w_{\pi})$
could reach large integer and half-integer values with the increase
of $J_{2}$. This provides us with another route to prepare non-Hermitian
Floquet phases with large integer/half-integer topological invariants. Besides, the existence
of Floquet phases with half-integer winding numbers also suggests
the necessity of introducing new topological invariants under the
OBC. In the second case, we present the topological phase diagram
of the system versus the hopping parameters $J_{1}$ and $J_{2}$
in Fig.~\ref{fig:PhsDiag3}, with the intracell hopping amplitude
and asymmetric coupling strength fixed at $\mu=0.4\pi$ and $\lambda=0.3$.
Despite the abundant non-Hermitian Floquet topological phases and
phase transitions that can be observed in Fig.~\ref{fig:PhsDiag3},
we also found phases with larger integer or half-integer winding numbers
$(w_{0},w_{\pi})$ at larger values of $J_{1}$ in certain ranges
of $J_{2}$, which again highlights the power of Floquet engineering
in generating non-Hermitian phases with large topological invariants.

Putting together, we conclude that the interplay between time-periodic
drivings and nonreciprocal effects could indeed yield rich non-Hermitian
Floquet topological phases with large integer and half-integer winding
numbers in the bulk. The winding numbers $(w_{0},w_{\pi})$ of a bulk non-Hermitian Floquet
phase can further be probed dynamically through the long-time averaged
spin textures introduced previously in Ref.~\cite{LWZNH3}. Meanwhile, the
observation of phases with half-integer winding numbers $(w_{0},w_{\pi})$
indicates the breakdown of the conventional bulk-edge correspondence
in Floquet systems and the emergence of NHSEs.

\subsection{Static spin textures and dynamic winding numbers\label{subsec:C}}
\begin{figure}
	\begin{centering}
		\includegraphics[scale=0.48]{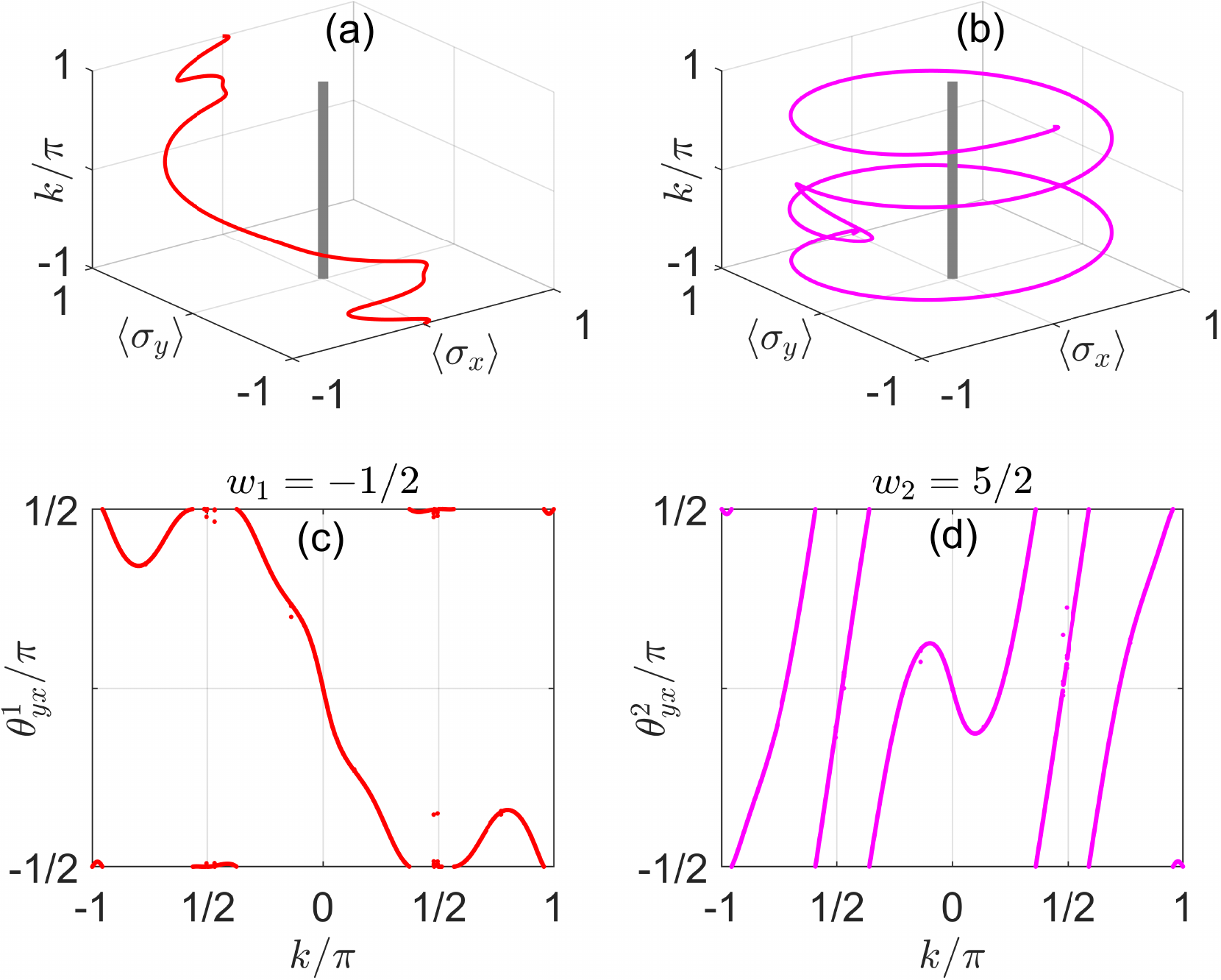}
		\par\end{centering}
	\caption{Static spin textures and dynamic winding angles of the periodically quenched NHSSH model in the times frame $\alpha=1$ [panels (a,c)] and $\alpha=2$ [panels (b,d)]. The system parameters are set as $(J_1,J_2,\mu,\lambda)=(1.4\pi,0.5\pi,0.4\pi,0.25)$, and the dynamic spin textures are averaged over $500$ driving peirods to generate the winding angles $\theta^{1,2}_{yx}$ in panels (c,d)~\cite{LWZNH3}. In panel (a) [(b)], the red (magenta) points denote the ends of the static spin vector $(\langle\sigma_x\rangle,\langle\sigma_y\rangle)$ in the first (second) time frame, and the grey solid line marks the origin of the $\langle\sigma_x\rangle$-$\langle\sigma_y\rangle$ plane. In panel (c) [(d)], the red (magenta) points correspond to the dynamic winding angles $\theta^{1}_{yx}$~($\theta^{2}_{yx}$) at different values of quasimomentum $k$ in the time frame $1$~($2$), which are calculated following Ref.~\cite{LWZNH3}. The winding numbers $(w_1,w_2)$ in both time frames are also shown explicitly in the corresponding panels.\label{fig:Spintext1}}
\end{figure}

In this section, we present the static spin textures and dynamic winding numbers~\cite{LWZNH3} of the periodically quenched NHSSH model for another situation in Fig.~\ref{fig:Spintext1}, where the panels (a) and (b) show the trajectors of spin vector $(\langle\sigma_x\rangle,\langle\sigma_y\rangle)$ versus the quasimomentum $k$ in the time frame $\alpha=1$ and $2$, respectively. The average $\langle\cdots\rangle$ is taken with respect to the right eigenvector $|\psi_{\alpha}^{+}(k)\rangle$ of $H_\alpha(k)$. The grey thick lines highlight the origin of the $\langle\sigma_x\rangle$-$\langle\sigma_y\rangle$ plane, which satisfy the equation $(\langle\sigma_x\rangle,\langle\sigma_y\rangle)=0$ at all $k\in[-\pi,\pi)$. In Figs.~\ref{fig:Spintext1}(a,b), we see that the geometric projections of $(\langle\sigma_x\rangle,\langle\sigma_y\rangle)$ onto the  $\langle\sigma_x\rangle$-$\langle\sigma_y\rangle$ plane contains an integer plus a half circle, which indicates the presence of half-integer winding numbers. For example, starting at $(\langle\sigma_x\rangle,\langle\sigma_y\rangle)=(0,1)$, the vector $(\langle\sigma_x\rangle,\langle\sigma_y\rangle)$  rotates counterclockwise around the origin over $4.5$ cycles, ending at $(\langle\sigma_x\rangle,\langle\sigma_y\rangle)=(0,-1)$ when $k$ sweeps from $-\pi$ to $\pi$, as shown in Fig.~\ref{fig:Spintext1}(b). These half-integer geometric windings are catched by the winding angles of dynamic spin textures $\theta^{1,2}_{yx}$, as shown in the panels (c,d) of Fig.~\ref{fig:Spintext1}~(see Ref.~\cite{LWZNH3} for the definition and calculation details of the dynamic winding angles), where the net increments of winding angles $\theta^{1,2}_{yx}$ across the first BZ are odd-integer multiples of $\pi$, yielding half-quantized integers after being divided by $2\pi$. In Ref.~\cite{LWZNH3}, it has been proved that $(\theta^{1}_{yx}/2\pi,\theta^{2}_{yx}/2\pi)$ are equal to the winding numbers $(w_1,w_2)$ defined in Eq.~(3) of the main text. Therefore, if $(\theta^{1}_{yx}+\theta^{2}_{yx})/(2\pi)$ or $(\theta^{1}_{yx}-\theta^{2}_{yx})/(2\pi)$ happens to be an odd integer, we will obtain a half-quantized invariant $w_0$ or $w_\pi$ according to Eq.~(4) of the main text. Thus, the half-integer quantizations of $(w_0,w_\pi)$ can also be dynamically extracted from the long-time averaged spin textures.

\subsection{Explicit forms of $H_{1,2}$ under the OBC\label{subsec:D}}
Under the OBC, the Hamiltonians of the system in the first and second halves of the driving period reads
\begin{alignat}{1}
H_{x}= & \frac{1}{2}\sum_{n}[\mu|n\rangle\langle n|+(J_{1}+\lambda)|n\rangle\langle n+1|\nonumber\\
& + (J_{1}-\lambda)|n+1\rangle\langle n|]\sigma_{x},\label{eq:H1OBC}\\
H_{y}= & \frac{i}{2}\sum_{n}[(\lambda-J_{2})|n\rangle\langle n+1|+(\lambda+J_{2})|n+1\rangle\langle n|]\sigma_{y},\label{eq:H2OBC}
\end{alignat}
with $H_{x,y}$ being the real-space representations of Hamiltonians $\sum_k h_{x}(k)\sigma_{x}$
and $\sum_k h_{y}(k)\sigma_{y}$ in Eq.~(1) of the main text, respectively.

\subsection{Open-boundary ${\cal Q}$-matrix\label{subsec:E}}
For a system described by a non-Hermitian Hamiltonian $H$, the eigenvalue problem can be studied in biorthogonal basis~\cite{BQM1}, in which the right and left eigenvectors satisfy the eigenvalue equations $H|\psi_{n}\rangle=E_{n}|\psi_{n}\rangle$ and $\langle\tilde{\psi}_{n}|H=\langle\tilde{\psi}_{n}|E_{n}$. Here $E_{n}\in{\mathbb C}$ is the eigenvalue of $H$ and $n=1,...,L$, with $L$ being the dimension of the system's Hilbert space. The eigenvectors $\{|\tilde{\psi}_{n}\rangle\}$ and $\{|\psi_{n}\rangle\}$ also satisfy the biorthogonal condition $\langle\tilde{\psi}_{m}|\psi_{n}\rangle=\delta_{mn}$ for $m,n=1,...,L$. If $H$ further has the chiral (sublattice) symmetry ${\cal S}$, i.e., ${\cal S}H{\cal S}=-H$, the right eigenvectors $(|\psi_{n}\rangle,{\cal S}|\psi_{n}\rangle)$ form a chiral symmetric partner with eigenvalues $(E_{n},-E_n)$. In this case, both the left and right eigenvectors of $H$ can be decomposed into two sets as $\{|\tilde{\psi}_{n}\rangle\}=\{|\tilde{\psi}_{n}^{+}\rangle,|\tilde{\psi}_{n}^{-}\rangle\}$ and $\{|\psi_{n}\rangle\}=\{|\psi_{n}^{+}\rangle,|\psi_{n}^{-}\rangle\}$, where $\{|\tilde{\psi}_{n}^{+}\rangle,|\psi_{n}^{+}\rangle\}$ and $\{|\tilde{\psi}_{n}^{-}\rangle,|\psi_{n}^{-}\rangle\}$ have the eigenvalues $\{E_{n}\}$ and $\{-E_{n}\}$. One can then construct an open-boundary ${\cal Q}$-matrix~\cite{NHSSH0} as ${\cal Q}=\sum_{n}(|\psi_{n}^{+}\rangle\langle\tilde{\psi}_{n}^{+}|-|\psi_{n}^{-}\rangle\langle\tilde{\psi}_{n}^{-}|)$, where the sum is taken over all bulk states of $H$. Physically, ${\cal Q}$ can be viewed as a biorthogonal flat-band projector, which is obtained by replacing the eigenvalues of $H$ by $+1$ ($-1$) if their real parts are positive (negative). The eigenstates with zero eigenvalues are excluded in the definition of ${\cal Q}$, as they correspond to the edge states of a chiral symmetric 1D lattice. In terms of ${\cal Q}$, an open-boundary winding number~(OBWN) can be introduced as $\nu=\frac{1}{L_{\rm B}}{\rm Tr}_{\rm B}({\cal S}{\cal Q}[{\cal Q},{\cal N}])$~\cite{NHSSH0,OBCWA1,OBCWA2,OBCWA3}. Here ${\cal N}$ denotes the position operator of unit cells. The system is decomposed into a bulk region and two edge regions around the left and right boundaries. The trace ${\rm Tr}_{\rm B}$ is taken over the bulk region, which contains $L_{\rm B}$ lattice sites. For a lattice of $L$ sites, the length of each edge region is $L_{\rm E}=(L-L_{\rm B})/2$. It has been demonstrated that $\nu$ correctly counts the number of zero-energy edge modes in 1D chiral-symmetric non-Hermitian lattices. Furthermore, $\nu$ is shown to be quantized and essentially equivalent to the non-Bloch winding number defined in the generalized Brillouin zone~\cite{NHSSH0}. Therefore, it could be employed to recover the bulk-edge correspondence in static 1D non-Hermitian systems.

In numerical calculations, the left and right eigenstates of $H$ can be obtained by writing $H=VDV^{-1}$, where $D$ is a diagonal matrix containing the eigenvalues of $H$, and the right (left) eigenvectors are given by the columns of $V$ {[}$(V^{-1})^{\dagger}${]}~\cite{NHSSH0}. The errors introduced by incorporating the zero-energy edge states in the evaluation of ${\cal Q}$ tends out to be negligible, so long as $L$ is much larger then the number of edge states in the system. In the meantime, $L_{\rm E}$ should be chosen sufficiently large to ensure that the trace ${\rm Tr}_{\rm B}$ only accounts the information of the bulk states~\cite{NHSSH0}.

\subsection{More examples of the Floquet spectrum and open-boundary winding numbers\label{subsec:F}}

\begin{figure}
	\begin{centering}
		\includegraphics[scale=0.48]{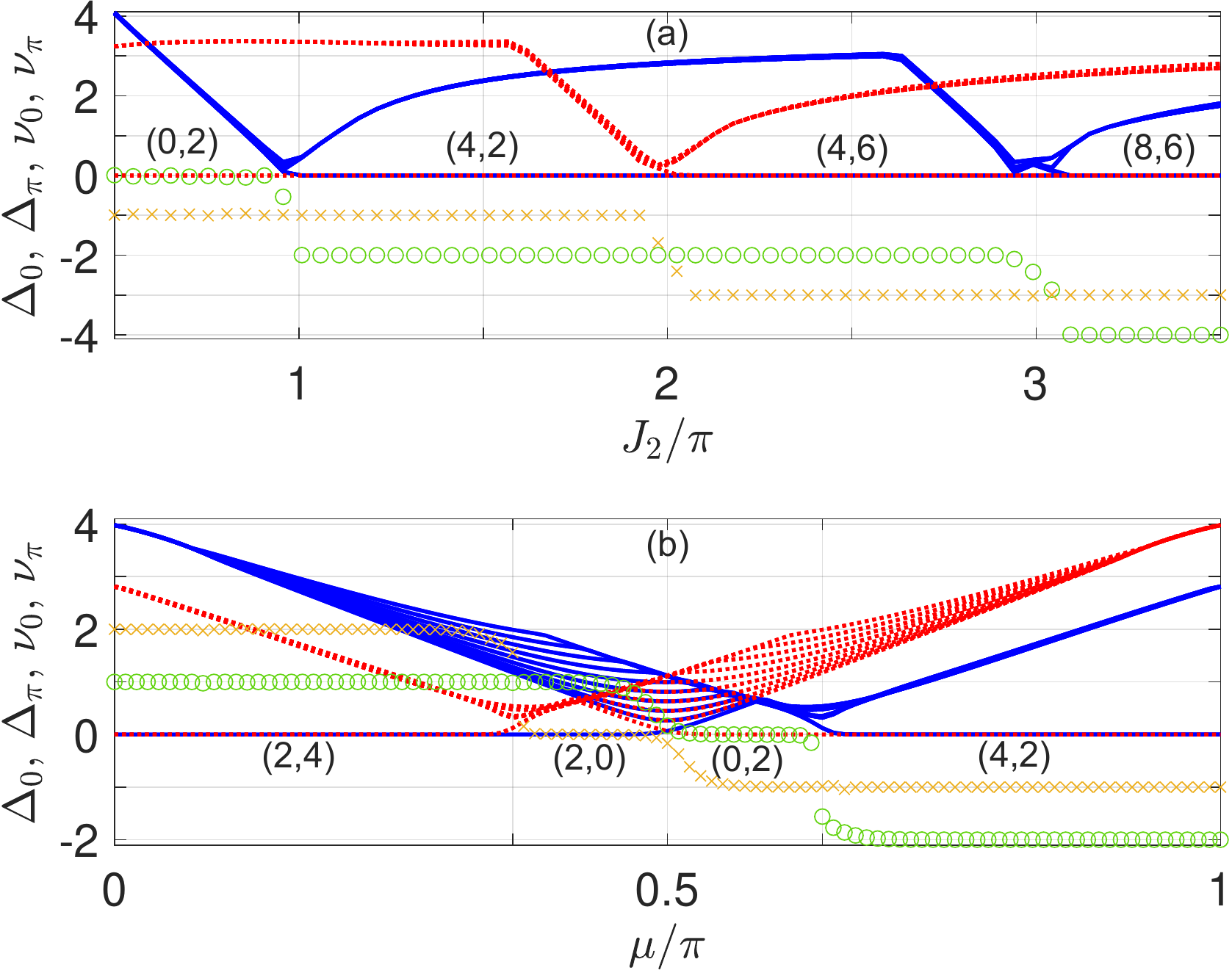}
		\par\end{centering}
	\caption{Gap functions $\Delta_{0}$ (blue solid lines), $\Delta_{\pi}$
		(red dotted lines), and open-boundary winding numbers $\nu_{0}$ (green
		circles), $\nu_{\pi}$ (yellow crosses) of the periodically quenched
		NHSSH model versus the hopping amplitude $J_{2}$ ($\mu$) in panel
		(a) {[}(b){]}. The system parameters are chosen as $(J_{1},\mu,\lambda,L,L_{\rm E})=(0.4\pi,\pi,0.25,240,90)$
		for panel (a) and $(J_{1},J_{2},\lambda,L,L_{\rm E})=(0.5\pi,1.5\pi,0.25,400,80)$
		for panel (b), where $L$ and $L_{\rm E}$ are the length of lattice and
		the number of lattice sites in the boundary region. For illustration
		purpose, only the first twenty smallest values of $(\Delta_{0},\Delta_{\pi})$
		are shown at each values of $J_{2}$ and $\mu$ in the two panels. Topological
		phase transitions happen around $J_{2}=(\pi,2\pi,3\pi)$ in panel
		(a) and $\mu=(0.36\pi,0.64\pi)$ in panel (b), which are highlighted
		by the ticks along the horizontal axis. A topological phase transition
		happens every time when $\Delta_{0}=0$ or $\Delta_{\pi}=0$, which
		is accompanied by the quantized jump of $\nu_{0}$ or $\nu_{\pi}$ by an integer.
		The numbers of zero and $\pi$ Floquet edge modes $(n_{0},n_{\pi})$
		for each non-Hermitian Floquet topological phase are also denoted
		along the horizontal axis in both panels.\label{fig:BBC2}}
\end{figure}
In this section, we give more numerical examples of the Floquet spectrum and OBWNs in order to verify the bulk-edge correspondence we discovered in the main text.
We plot the open-boundary spectral gap functions
$(\Delta_{0},\Delta_{\pi})$ and winding numbers $(\nu_{0},\nu_{\pi})$
with respect to the hopping amplitudes $J_{2}$ and $\mu$ in panels
(a) and (b) of Fig.~\ref{fig:BBC2}, respectively. In both panels,
we obtain the correct relationship between the values of bulk invariants
$(\nu_{0},\nu_{\pi})$ and the numbers of edge modes $(n_{0},n_{\pi})$,
as described by $(n_{0},n_{\pi})=2(|\nu_{0}|,|\nu_{\pi}|)$ for all non-Hermitian Floquet
topological phases. These observations further confirm the universality
of our approach in characterizing the topological phases and bulk-edge
correspondence in 1D chiral-symmetric non-Hermitian Floquet systems,
even in the concomitance NHSEs. Besides, our approach does not rely
on the explicit calculation of a generalized BZ. This could be an
advantage when the system under investigation contains long-range hoppings, disorder, nonlinear
effects or many-body interactions, for which the single-particle BZ
is either difficult to obtain or not well-defined. For example, we have numerically verified that the open-boundary winding numbers could faithfully capture most of the Floquet topological phases with NHSEs in the non-perturbative regions of the model studied in Ref.~\cite{FNHSE1}, in which the non-Bloch BZ cannot be constructed accurately.
However, in the numerical calculations
of $(\nu_{0},\nu_{\pi})$, the most appropriate size of the bulk region
$L_{\rm B}$ for a given lattice of length $L$ cannot be determined in
advance, and some trials and errors are needed in order to find the optimized
length of the bulk region $L_{\rm B}$. The full resolution of this issue will be deferred to future work. Putting together, we conclude that the periodically
quenched NHSSH model could indeed possess rich non-Hermitian topological
phases and multiple Floquet edge modes at zero and $\pi$ quasienergies,
which are coexisting with the NHSEs. The OBWNs
$(\nu_{0},\nu_{\pi})$ could provide us with an efficient characterization of
the non-Hermitian Floquet topological phases and bulk-edge correspondence
in the system.

\subsection{Non-Hermitian Floquet skin modes\label{subsec:G}}
\begin{figure}
\begin{centering}
\includegraphics[scale=0.41, trim=0cm 6.9cm 0cm 7.4cm, clip=true]{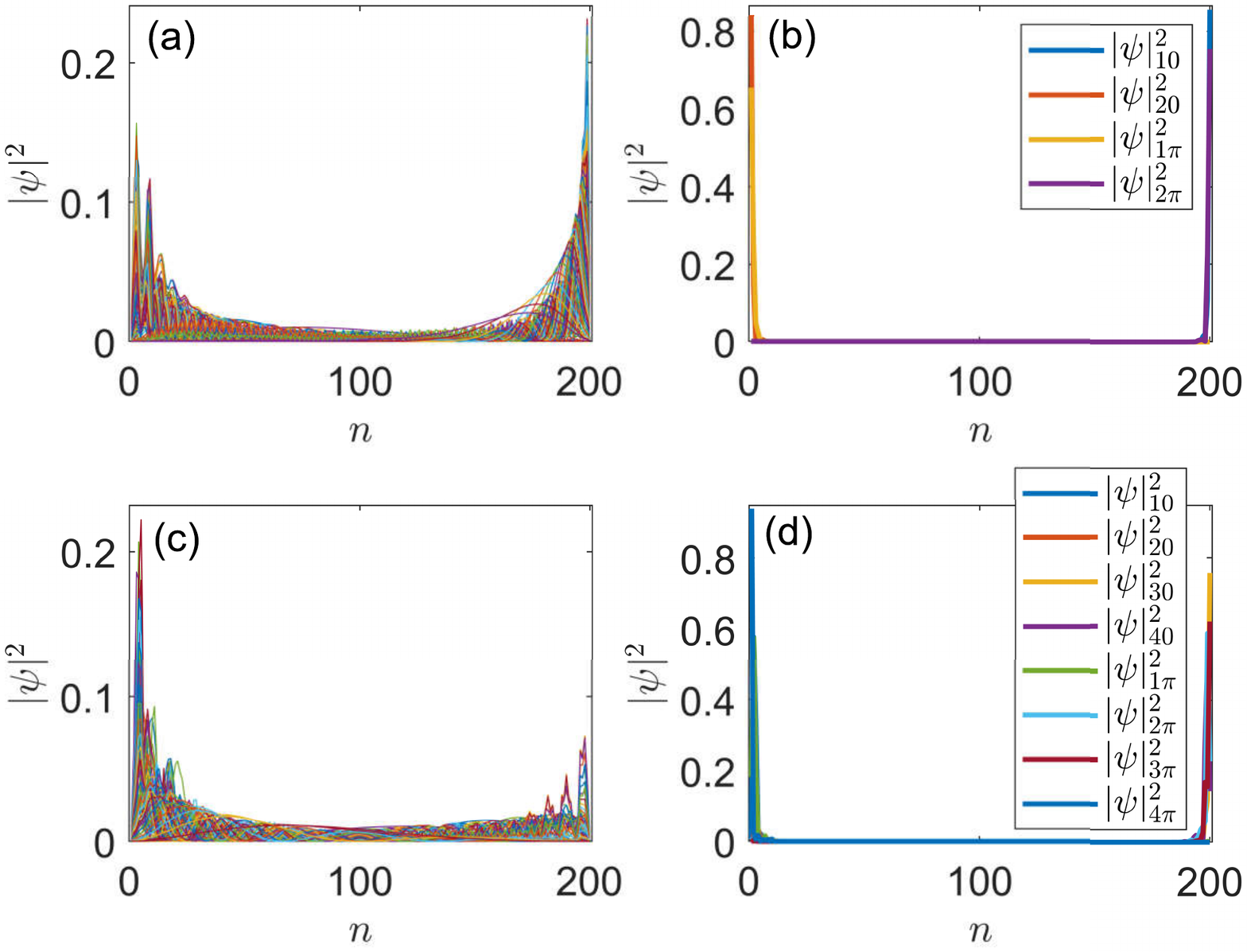}
\par\end{centering}
\caption{Probability distributions of bulk and edge states of the periodically quenched NHSSH model studied in the main text. System parameters are $(J_{1},J_{2},\mu,\lambda)=(\pi,0.5\pi,0.4\pi,0.25)$ for panels (a,b) and $(J_{1},J_{2},\mu,\lambda)=(2\pi,0.5\pi,0.4\pi,0.25)$ for panels (c,d). $n$ is the unit-cell index. The number of unit cells is $200$. In panels (a,c), NHSEs are demonstrated by the accumulation of bulk states around both edges of the lattice. In panel (b) {[}(d){]}, there are one pair (two pairs) of Floquet edge modes at both eigenphases zero and $\pi$, coexisting with the bulk Floquet skin modes.\label{fig:BulkEdgeState}}
\end{figure}

To demonstrate the coexistence of NHSEs and topological edge modes in the periodically quenched NHSSH model, we show the profiles of bulk and edge states of our system in Fig.~\ref{fig:BulkEdgeState}. We observe that many bulk states are piling up around both of the boundaries, exhibiting the NHSEs. Meanwhile, edge modes with zero and $\pi$ eigenphases are found at the left and right edges of the system, coexisting with the Floquet non-Hermitian skin modes. Notably, even though the probability distributions of skin and edge modes are both close to the boundaries, their quasienergies are well-separated on the complex plane, as can be inferred from the spectrum gap functions in Fig.~4 of the main text.

\end{document}